\documentstyle[aps,prb,twocolumn,epsf,floats]{revtex}
\begin{document}
\draft
\wideabs{
\title{
   Interaction and Particle--Hole Symmetry of Laughlin Quasiparticles}
\author{
   Arkadiusz W\'ojs}
\address{
   Department of Physics, University of Tennessee, 
   Knoxville, Tennessee 37996, and \\
   Institute of Physics, Wroclaw University of Technology,
   50-370 Wroclaw, Poland}
\maketitle
\begin{abstract}
   The pseudopotentials describing interaction of Laughlin 
   quasielectrons (QE) and quasiholes (QH) in an infinite 
   fractional quantum Hall system are studied.
   The QE and QH pseudopotentials are similar which suggests
   the (approximate) particle--hole symmetry recovered in the 
   thermodynamical limit.
   The problem of the hypothetical symmetry-breaking QE hard-core 
   repulsion is resolved by the estimate that the ``forbidden'' 
   QE pair state has too high energy and is unstable.
   Strong oscillations of the QE and QH pseudopotentials 
   persist in an infinite system, and the analogous QE and QH 
   pair states with small relative angular momentum and nearly 
   vanishing interaction energy are predicted.
\end{abstract}
}

An important element in our understanding of the incompressibile-fluid 
ground states\cite{laughlin,tsui,prange} formed in a two-dimensional 
electron gas (2DEG) in high magnetic fields has been the identification 
of Laughlin correlations\cite{laughlin} in a partially filled lowest 
Landau level (LL).
These correlations can be defined\cite{parentage,correlations} as 
a tendency to avoid pair eigenstates with the largest repulsion 
(smallest relative pair angular momentum ${\cal R}$) in the low-energy 
many-body states.
The incompressibility results at a series of filling factors (number 
of particles divided by the number of states) $\nu=(2p+1)^{-1}$ at 
which the $p$ leading pair states at ${\cal R}=1$, 3, \dots, $2p-1$ 
are completely avoided in the non-degenerate (Laughlin) ground state, 
but not in any of the excited states.

Each Laughlin-correlated state can be understood in terms of two types
of quasiparticles (QP): quasielectrons (QE) and quasiholes (QH), moving 
in an underlying Laughlin ground state (``reference'' or ``vacuum'' 
state).
The QP's are the elementary excitations of the Laughlin fluid and 
correspond to an excessive (QH) or missing (QE) single-particle state, 
compared to an exact $\nu=(2p+1)^{-1}$ filling.
They have finite size and (fractional) electric charge of 
$\pm(2p+1)^{-1}e$, and thus (in analogy to LL's of electrons) the
single-QP spectrum in a magnetic field is degenerate at a finite 
energy denoted as $\varepsilon_{\rm QP}$.
For the QP's at a complex coordinate $z=0$, their wavefunctions are 
obtained by applying the prefactors $\prod_k \partial/\partial z_k$ 
(QE) and $\prod_k z_k$ (QH) to the Laughlin wavefunction $\Phi_{2p+1}=
\prod_{i<j}(z_i-z_j)^{2p+1}$.

Partially filled lowest LL is not the only many-body system with Laughlin 
correlations, which generally occur when the single-particle Hilbert 
space is degenerate and the two-body interaction is repulsive and has 
short range.\cite{parentage,correlations}
Among other Laughlin-correlated systems are a two-component system 
of electrons and charged excitons ($X^-$, two electrons bound to a 
valence hole) formed in an electron--hole plasma in a magnetic field,
\cite{xminus,2comp} or a system of (bosonic) electron pairs formed 
near the half-filling of the first excited LL\cite{correlations} 
(Moore--Read\cite{moore} state at $\nu={5\over2}$).

Due to their LL-like macroscopic degeneracy and the Coulomb nature of
their interaction, Laughlin correlations can also be expected in a 
system of Laughlin QP's.
The concept of Laughlin ground states formed by Laughlin QP's gave 
rise to Haldane's hierarchy\cite{haldane-hierarchy} of incompressible 
``daughter'' states at $\nu_{\rm QP}=(2p_{\rm QP}+1)^{-1}$, in addition 
to the ``parent'' states at $\nu=(2p+1)^{-1}$.
For example, the incompressible $\nu={2\over7}$ state can be viewed 
as the $\nu_{\rm QH}={1\over3}$ state of QH's in the parent 
$\nu={1\over3}$ state of electrons.

The criterion for the ``short range'' of the two-body repulsion that 
causes Laughlin correlations is expressed in terms of the interaction 
pseudopotential $V({\cal R})$, defined\cite{parentage,haldane-pseudo}
as the pair interaction energy $V$ as a function of ${\cal R}$.
Therefore, the knowledge of $V({\cal R})$ is necessary to predict 
the type of correlations (and possible incompressibility) in a given 
many-body system.

In this note we continue our earlier study\cite{hierarchy} of 
interactions between Laughlin QP's.
The QE and QH interaction pseudopotentials are calculated for the 
Laughlin $\nu={1\over3}$ and ${1\over5}$ states of up to 8 and 12 
electrons on a Haldane sphere,\cite{haldane-hierarchy} respectively, 
and extrapolated to an infinite planar system.
Our results lead to the following two main conclusions:

(i) 
Opposite to what seemed to follow from finite-size 
calculations,\cite{hierarchy,yi} the sign and magnitude of the 
pseudopotential coefficients calculated for an infinite plane agree 
with the expectation that, being charge excitations, the QP's of the 
same type must repel and not attract one another.
However, the oscillations in the QP charge density cause oscillations 
in $V({\cal R})$, and the QP pair states with small ${\cal R}$ (small 
radius) and nearly vanishing interaction energy are predicted.
The vanishing of repulsion in these states rules out incompressibility 
of such hypothetical\cite{hierarchy} ground states in Haldane's 
hierarchy as $\nu={6\over17}$ or ${6\over19}$, and limits the family 
of valid hierarchy states to the (experimentally observed) Jain's 
sequence\cite{jain} at $\nu=n(2pn\pm1)^{-1}$.
This vanishing is also essential for the stability of fractionally 
charged excitons\cite{fcx} $h$QE$_n$ ($n$ QE's bound to a valence 
hole) observed\cite{heiman} in photoluminescence of the 2DEG.

(ii)
From the similarity of QE and QH pseudopotentials we conclude that, 
in agreement with Haldane's intuitive picture of QP's placed ``between'' 
the electrons,\cite{haldane-hierarchy} no asymmetry between QE and QH 
Hilbert spaces (angular momenta) exists that should require introduction 
of a phenomenological hard-core QE--QE repulsion\cite{he} to predict 
the correct number of many-QE states in numerical energy spectra.
Instead, the repulsion energy of the ``forbidden'' QE pair state is 
finite but higher than that of a corresponding QH pair state.
It even exceeds the Laughlin gap $\Delta=\varepsilon_{\rm QE}+
\varepsilon_{\rm QH}$ to create an additional QE--QH pair and makes 
this QE pair state unstable (and pushes it into the 3QE$+$QH continuum).
This instability explains why Jain's composite fermion (CF) picture
\cite{jain} correctly predicts the lowest-energy bands of states, 
despite the asymmetry of QE and QH LL's introduced by an (unphysical) 
effective magnetic field.
Also, the similarity of the QE and QH pair states and energies 
precludes qualitatively different response of a Laughlin-correlated
2DEG to a positively and negatively charged perturbation.\cite{fcx}

The knowledge of pseudopotentials defining interactions of Laughlin 
QP's is essential in Haldane's hierarchy\cite{haldane-hierarchy} 
of the fractional quantum Hall effect,\cite{laughlin,tsui,prange} 
in which they determine those of Laughlin fillings at which the QP's 
form (daughter) Laughlin incompressible states of their own.
Although they are to a large extent equivalent, Haldane's hierarchy 
differs from Jain's CF picture in the ``symmetric'' description of 
the two types of QP's.
Haldane's elegant argument\cite{haldane-hierarchy} that both QE and 
QH excitations are bosons placed ``between'' the $N$ (effectively 
one-dimensional) electrons yields equal numbers of possible QE and 
QH states, $\tilde{g}_{\rm QE}=\tilde{g}_{\rm QH}=N+1$ (tildes mean 
bosons), which on a sphere correspond to equal single-particle angular 
momenta, $\tilde{l}_{\rm QE}=\tilde{l}_{\rm QH}={1\over2}N$ 
(because $\tilde{g}=2\tilde{l}+1$; the lowest LL on a Haldane sphere 
is an angular momentum shell of $l=S$, half the strength of Dirac's 
magnetic monopole in the center\cite{parentage,haldane-hierarchy,chen}).

In a system of $n$ QP's, a mean-field Chern--Simons transformation 
(MFCST)\cite{canright,wilczek,statistics} can further be used to 
convert such bosonic QP's to more convenient fermions with 
$g=\tilde{g}+(n-1)$ yielding $l=\tilde{l}+{1\over2}(n-1)$.
However, for QE's this value of $l$ seemed to predict incorrect 
number of low-energy states in the numerical energy spectra unless
the pair state at the maximum angular momentum $L_{\rm max}=2l-1$ 
was forbidden.\cite{he}
On a sphere, the relation between $L=|{\bf l}_1+{\bf l}_2|$ and 
${\cal R}$ is $L=2l-{\cal R}$, and thus the exlusion of the pair
state at $L_{\rm max}$ is equivalent to a hypothetical hard-core 
repulsion, $V_{\rm QE}(1)=\infty$.

Such interaction hard-core can be formally removed by an 
appropriate redefinition of the single-particle Hilbert space.
This is accomplished by a fermion-to-fermion MFCST
\cite{parentage,hierarchy,lopez,halperin} which replaces $g$ 
by $g^*=g-2(n-1)$, and $l={1\over2}N+{1\over2}(n-1)$ by $l^*=
{1\over2}N-{1\over2}(n-1)$.
By ``elimination'' we mean that the angular momenta $L_{n{\rm QE}}$ 
of states containing $n$ QE's can be obtained by simple and 
unrestricted addition of $n$ individual angular momentum vectors 
${\bf l}_{\rm QE}^*$ followed by antisymmetrization (QE's are treated 
as indistinguishable fermions), just as $L_{n{\rm QH}}$ could be 
obtained by antisymmetric combination of $n$ vectors ${\bf l}_{\rm QH}$.
Although they seem to agree with the ``numerical experiments,'' 
\cite{parentage,hierarchy,he} no explanation exists for a hard-core 
in the QE--QE repulsion (and its absence in $V_{\rm QH}$) or the 
resulting asymmetry between $l_{\rm QH}$ and $l_{\rm QE}^*$.

This asymmetry is inherent in Jain's CF picture,\cite{jain} in which 
QE's and QH's are converted into particles and vacances in different 
CF LL's whose (different) angular momenta are equal to $l_{\rm QE}^*$ 
and $l_{\rm QH}$, respectively.
However, the effective magnetic field leading to the correct 
values of $g_{\rm QE}^*$ and $g_{\rm QH}$ in the CF picture does 
not physically exist.
While for the QH states the effective field is one of possible 
physical realizations of the MFCST describing Laughlin correlations 
(the avoidance of most strongly repulsive pair states) in the 
underlying electron system,\cite{parentage} no explanation for 
$g_{\rm QE}^*$ being smaller than $g_{\rm QH}$ is possible within 
the CF model itself.

\begin{figure}[t]
\epsfxsize=3.35in
\epsffile{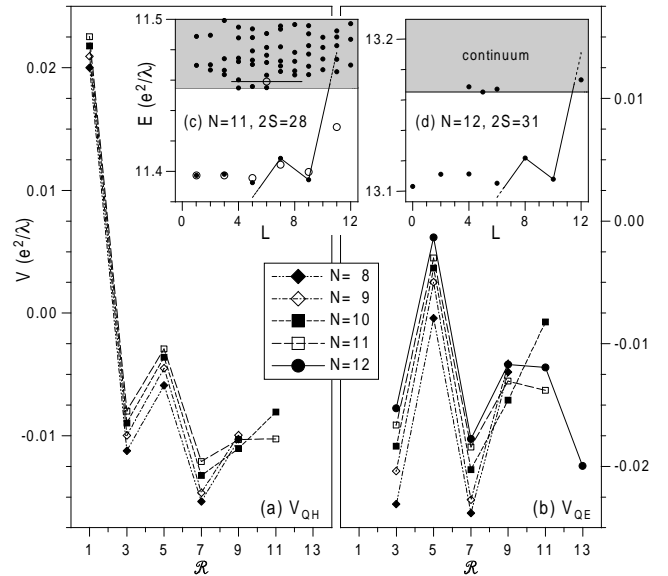}
\caption{
   The comparison of quasihole (a) and quasielectron (b) 
   pseudopotentials $V({\cal R})$ calculated at $\nu={1\over3}$
   in $N$-electron systems on a Haldane sphere.
   Insets: 
   The comparison of 11-electron (c) and 12-electron (d) energy 
   spectra in which the lowest-energy band contains two 
   quasielectrons.
   In (c), open circles show the (shifted in energy) 11-electron 
   spectrum of two quasiholes.
}
\label{fig1}
\end{figure}

To resolve this puzzle we have examined the QE and QH pseudopotentials 
calculated for the systems of $N\le12$ electrons at $\nu={1\over3}$ 
and ${1\over5}$.
In Fig.~\ref{fig1} we compare $V_{\rm QE}$ (a) and $V_{\rm QH}$ 
(b) obtained at $\nu={1\over3}$ for different values of $N$.
In both frames, ${\cal R}=2l-L$, with $l_{\rm QE}=l_{\rm QH}=
{1\over2}(N+1)$.
To obtain the values of $V$, the energies of the Laughlin ground 
state and of the two QP's are subtracted from the energies of the 
appropriate QP pair states\cite{hierarchy,yi} (such as the QE pair
states for $N=11$ and 12 shown in the insets).
The energy is measured in the units of $e^2/\lambda$, and $\lambda$ 
is the magnetic length.

In the limit of $N\rightarrow\infty$, the sphere radius $R\sim\sqrt{N}$ 
diverges and the numerical values of $V({\cal R})$ converge to those 
describing an infinite 2DEG on a plane.
In this (planar) geometry, ${\cal R}$ is the usual relative pair angular 
momentum.
Remarkably, when ${\cal R}_{\rm QE}$ is defined as $2l_{\rm QE}-L$ 
rather than $2l_{\rm QE}^*-L$, the QE and QH pseudopotentials become 
quite similar.
The main difference is the obvious lack of the ${\cal R}_{\rm QE}=1$ 
state and stronger oscillations in the $V_{\rm QE}({\cal R})$, but
the maximum at ${\cal R}=5$ and the minima at ${\cal R}=3$ and 7 are 
common for both $V_{\rm QE}$ and $V_{\rm QH}$.
The same structure occurs also for the QP'as in the $\nu={1\over5}$ 
state.\cite{hierarchy}
Most unexpected in Fig.~\ref{fig1} are the negative signs of 
$V_{\rm QP}$.
The only positive pseudopotential coefficient is $V_{\rm QH}(1)$,
which might indicate that, despite QP's being charge excitations, 
both QE--QE and QH--QH interactions are generally attractive.

\begin{figure}[t]
\epsfxsize=3.35in
\epsffile{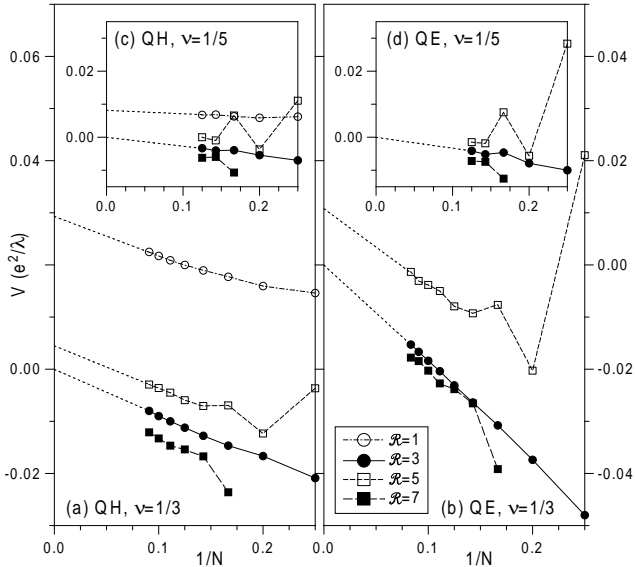}
\caption{
   The leading quasihole (ac) and quasielectron (bd) pseudopotential 
   coefficients $V({\cal R})$ calculated at $\nu={1\over3}$ (ab) and
   $\nu={1\over5}$ (cd) in $N$-electron systems on a Haldane sphere,
   plotted as a function of $N^{-1}$.
   Thin dotted lines show extrapolation to $N\rightarrow\infty$.
}
\label{fig2}
\end{figure}

In Fig.~\ref{fig2} we plot a few leading pseudopotential coefficients 
(those at the smallest values of ${\cal R}$) $V_{\rm QH}$ and 
$V_{\rm QE}$ at $\nu={1\over3}$ and ${1\over5}$ as a function of 
$N^{-1}$.
Clearly, the corresponding coefficients of all four pseudopotentials
behave similarly which confirms the correct use of $l_{\rm QE}$ rather
than $l_{\rm QE}^*$ in the definition of ${\cal R}_{\rm QE}$.
It is also clear that all coefficients $V$ increase with increasing 
$N$ (although at a different rate for QE's and QH's) and it seems that 
none of them will remain negative in the $N\rightarrow\infty$ limit.
In attempt to estimate the magnitude of $V$ in this limit we have
drawn straight linest that approximately extrapolate our data for
some of the coefficients.
Most noteworthy values are:
$V_{{\rm QH},\,\nu=1/3}(1)\approx0.03\,e^2/\lambda$ being about three 
times larger than $V_{{\rm QH},\,\nu=1/5}(1)$ as expected from the 
comparison of interacting charges (${1\over3}e$ and ${1\over5}e$, 
respectively), the $V(3)$ coefficients (seemingly) vanishing in all 
four plots, and $V_{{\rm QH},\,\nu=1/3}(5)\approx0.005\,e^2/\lambda$ 
being about twice smaller than $V_{{\rm QE},\,\nu=1/3}(5)$.

The predicted small value of $V(3)$ and of some other leading 
coefficients is by itself quite interesting, although it can be 
understood from the fact that QP's are more complicated objects 
than electrons, and the oscillations in $V_{\rm QP}({\cal R})$ 
reflect the oscillations in their more complicated charge density 
profile (similar oscillations occur in the electron pseudopotentials 
in higher LL's).
The consequences of this fact are even more important.

First, from a general criterion\cite{parentage,correlations} for 
Laughlin correlations at $\nu\approx(2p+1)^{-1}$ (defined as the 
avoiding of pair states with ${\cal R}<2p+1$ in the low-energy 
many-body states) in a system interacting through a pseudopotential 
$V({\cal R})$ we find that the QP's of the parent Laughlin state 
of electrons form Laughlin states of their own only 
at $\nu_{\rm QP}={1\over3}$.
These states and their $\nu_{\rm QE}={1\over3}$ daughters exhaust
Jain's $\nu=n(2pn\pm1)^{-1}$ sequence.
No other incompressible daughter states occur in the hierarchy, 
including the (ruled out earlier\cite{hierarchy}) $\nu={4\over11}$ 
or ${4\over13}$ states or the hypothetical\cite{hierarchy} 
$\nu={6\over17}$ or ${6\over19}$ states.
Despite all the differences between Haldane's hierarchy and Jain's 
CF model, our conclusion makes their predictions of the 
incompressibility at a given $\nu$ completely equivalent.

Second, the (near) vanishing of $V_{\rm QE}(3)$ explains the stability
of the $h$QE$_2$ complex\cite{fcx} in the 2DEG interacting with an 
(optically injected) valence hole.
Being the most strongly bound and the only radiative state of all 
``fractionally charged excitons'' $h$QE$_n$, the $h$QE$_2$ is most 
likely the complex observed\cite{heiman} in the PL spectra of the 
2DEG at $\nu>{1\over3}$.

Third, since $V_{\rm QE}(5)$ is about twice larger than $V_{\rm QH}(5)$,
it is also plausible that $V_{\rm QE}(1)$ could be much larger than 
$V_{\rm QH}(1)$, so that the ${\cal R}_{\rm QE}=1$ state would fall 
in the continuum and could not be identified in the energy spectra.
In Fig.~\ref{fig1}(c), on top of the 11-electron spectrum at Dirac's 
monopole strength (the number of magnetic flux quanta piercing the 
Haldane sphere\cite{haldane-hierarchy,parentage,chen}) $2S=28$, 
marked with full dots, in which the lowest-energy states contain 
two QE's at $\nu={1\over3}$, with open circles we have marked 
another spectrum calculated for the same $N=11$ but at $2S=32$, 
whose lowest-energy band contains two QH's.
The second spectrum is vertically shifted so that the energies of 
the QE and QH pair states coincide at $L=1$ (i.e.\ at ${\cal R}=11$) 
at which $V_{\rm QE}$ and $V_{\rm QH}$ are both negligible), but the 
energy units ($e^2/\lambda$) are the same.
Since the Laughlin gap $\Delta$ to the continuum of states with 
additional QE--QH pairs involves the sum of QE and QH energies, 
it is roughly the same in both spectra.
However, the minima and maxima in $V_{\rm QE}({\cal R})$ are 
stronger than those in $V_{\rm QH}({\cal R})$, and the difference 
$|V_{\rm QE}-V_{\rm QH}|$ increases at larger $L$.
While it is hardly possible to rescale $V_{\rm QH}$ so as to reproduce 
$V_{\rm QE}$ at $L\le9$ and convincingly predict its value at $L=11$ 
(${\cal R}_{\rm QE}=11$), it seems likely that $V_{\rm QE}(11)$ is 
indeed larger than $\Delta$, which would explain the absence of the 
${\cal R}_{\rm QE}=11$ state below the continuum.
An example of such ``rescaling'' procedure is shown in Fig.~\ref{fig1}(c) 
with the line obtained by stretching $V_{\rm QH}$ so that it crosses 
$V_{\rm QE}(5)$ and $V_{\rm QE}(3)$ at $L=7$ and 9, respectively.
Similar lines are shown in Fig.~\ref{fig1}(d) for the 12-electron 
spectrum corresponding to two QE's in the lowest band ($2S=31$).
Certainly, this procedure, based on the assumption that $V_{\rm QE}(3)$ 
and $V_{\rm QH}(3)$ are small and that $V(1)$ is proportional to $V(5)$,
is not accurate.
Nevertheless, having in mind the similarities of $V_{\rm QE}$ and 
$V_{\rm QH}$ in Figs.~\ref{fig1} and \ref{fig2}, and in the absence
of any physical reason why the ${\cal R}_{\rm QE}=1$ state might not 
exist while the ${\cal R}_{\rm QH}=1$ state does, we believe that it
is more reasonable to assume that $V_{\rm QE}(1)$ is finite, although
larger than $\Delta$.
The fact that the ${\cal R}_{\rm QE}=1$ state is pushed into the 
3QE$+$QH continuum simply means that it is unstable toward 
spontaneous creation of a low-energy QE--QH pair with finite angular 
momentum (magneto-roton).

The assumption that $\Delta<V_{\rm QE}(1)<\infty$ restores the elegant 
symmetry of Haldane's picture of QP's ``placed'' between electrons.
\cite{haldane-hierarchy}
It replaces the problem of explaining the QE hard-core\cite{he} by 
a question of why $V_{\rm QE}$ is larger than $V_{\rm QH}$ at short 
distance (e.g., at ${\cal R}=1$ and 5; see Fig.~\ref{fig1}).
But the fact that $V_{\rm QE}$ and $V_{\rm QH}$ are not equal at short 
distance is by no means surprising since the QE and QH have different 
wavefunctions.

In conclusion, we have calculated the pseudopotentials 
$V_{{\rm QP},\,\nu}({\cal R})$ describing interaction of QE's and 
QH's in Laughlin $\nu=(2p+1)^{-1}$ states of an infinite 2DEG.
These pseudopotentials are all similar, showing strong repulsion at 
${\cal R}=1$ and 5, and virtually no interaction at ${\cal R}=3$.
The unexpected QE--QE and QH--QH attraction which results in  
few-electron calculations disappears in the limit of an infinite
system.
Because the QP charge at $\nu=(2p+1)^{-1}$ decreases with increasing 
$p$, the QP interaction at $\nu={1\over3}$ is stronger than at 
$\nu={1\over5}$.
Because of different QE and QH wavefunctions, $V_{\rm QE}$ is larger 
than $V_{\rm QH}$ at small ${\cal R}$ (short distance).
The coefficient $V_{\rm QE}(1)$ exceeds the Laughlin gap $\Delta$ to 
create an additional QE--QH pair, which makes the QE pair state at 
${\cal R}=1$ unstable.
This instability, rather than a mysterious QE hard-core or an inherent 
asymmetry between the QE and QH angular momenta, is the reason for the 
overcounting of few-QE states when, following Haldane, QE's are treated 
as bosons with $\tilde{l}={1\over2}N$.
In particular, it explains the absence of the $L=N$ multiplet in the 
low-energy band of states in the $N$-electron numerical spectra at the 
values of $2S=(2p+1)(N-1)-2$, corresponding to two QE's in the Laughlin 
$\nu=(2p+1)^{-1}$ state.
The (near) vanishing of $V_{\rm QP}(3)$ is the reason why no hierarchy
states other than those from Jain's $\nu=n(2pn\pm1)^{-1}$ sequence are
stable.
It is also the reason for the strong binding of the fractionally charged 
exciton $h$QE$_2$.

The gratefully acknowledges discussions with John J. Quinn, Pawel 
Hawrylak, Lucjan Jacak, and Izabela Szlufarska, and support from 
the Polish State Committee for Scientific Research (KBN) Grant 
No.\ 2P03B11118.

\end{document}